# The GRAVITY+ Project: Towards All-sky, Faint-Science, High-Contrast Near-Infrared Interferometry at the VLTI

GRAVITY+ Collaboration

Roberto Abuter[8]
Patricio Alarcon[8]
Fatme Allouche[19]
Antonio Amorim[6, 11]
Christophe Bailet[19]
Helen Bedigan[8]
Anthony Berdeu[2]
Jean-Philippe Berger[5]
Philippe Berio[19]
Azzurra Bigioli[16]
Richard Blaho[8]
Olivier Boebion[19]
Marie-Lena Bolzer[1, 12]
Henri Bonnet[8]
Guillaume Bourdarot[1]
Pierre Bourget[8]
Wolfgang Brandner[3]
Cesar Cardenas[8]
Ralf Conzelmann[8]
Mauro Comin[8]
Yann Clénet[2]
Benjamin Courtney-Barrer[8, 18]
Yigit Dallilar[1]
Ric Davies[1]
Denis Defrère[16]
Alain Delboulbé[5]
Françoise Delplancke-Ströbele[8]
Roderick Dembet[2]
Tim de Zeeuw[9, 1]
Antonia Drescher[1]
Andreas Eckart[4, 13]
Clemence Édouard[2]
Frank Eisenhauer[1]
Maximilian Fabricius[1]
Helmut Feuchtgruber[1]
Gert Finger[1]
Natascha M. Förster Schreiber[1]
Eloy Fuenteseca[8]
Enrique Garcia[8]
Paulo Garcia[7, 11]
Feng Gao[14,1]
Eric Gendron[2]
Reinhard Genzel[1, 10]
Juan Pablo Gil[8]
Stefan Gillessen[1]
Tiago Gomes[7, 11]
Frédéric Gonté[8]
Carole Gouvret[19]
Patricia Guajardo[8]
Ivan Guidolin[8]
Sylvain Guieu[5]
Ronald Guzmann[8]
Wolfgang Hackenberg[8]
Nicolas Haddad[8]
Michael Hartl[1]
Xavier Haubois[8]
Frank Haußmann[1]
Gernot Heißel[2]
Thomas Henning[3]
Stefan Hippler[3]
Sebastian Hönig[15]
Matthew Horrobin[4]
Norbert Hubin[8]
Estelle Jacqmart[19]
Laurent Jocou[5]
Andreas Kaufer[8]
Pierre Kervella[2]
Jean-Paul Kirchbauer[8]
Johan Kolb[8]
Heidi Korhonen[8]
Laura Kreidberg[3]
Peter Krempl[8]
Sylvestre Lacour[2, 8]
Stephane Lagarde[19]
Olivier Lai[19]
Vincent Lapeyrère[2]
Romain Laugier[16]
Jean-Baptiste Le Bouquin[5]
James Leftley[19]
Pierre Léna[2]
Steffan Lewis[8]
Dieter Lutz[1]
Yves Magnard[5]
Felix Mang[1, 12]
Aurelie Marcotto[19]
Didier Maurel[5]
Antoine Mérand[8]
Florentin Millour[19]
Nikhil More[1]
Hugo Nowacki[5]
Matthias Nowak[17]
Sylvain Oberti[8]
Francisco Olivares[8]
Thomas Ott[1]
Laurent Pallanca[8]
Thibaut Paumard[2]
Karine Perraut[5]
Guy Perrin[2]
Romain Petrov[19]
Oliver Pfuhl[8]
Nicolas Pourré[5]
Sebastian Rabien[1]
Christian Rau[1]
Miguel Riquelme[8]
Sylvie Robbe-Dubois[19]
Sylvain Rochat[5]
Muhammad Salman[16]
Malte Scherbarth[8]
Markus Schöller[8]
Joseph Schubert[1]
Nicolas Schuhler[8]
Jinyi Shangguan[1]
Pavel Shchekaturov[8]
Taro Shimizu[1]
Silvia Scheithauer[3]
Arnaud Sevin[2]
Christian Soenke[8]
Ferreol Soulez[20]
Alain Spang[19]
Eric Stadler[5]
Christian Straubmeier[4]
Eckhard Sturm[1]
Calvin Sykes[15]
Linda Tacconi[1]
Helmut Tischer[8]
Konrad Tristram[8]
Frédéric Vincent[2]
Sebastiano von Fellenberg[13, 1]
Sinem Uysal[1]
Felix Widmann[1]
Ekkehard Wieprecht[1]
Erich Wiezorrek[1]
Julien Woillez[8]
Şenol Yazıcı[1]
Gérard Zins[8]

[1] Max Planck Institute for Extraterrestrial Physics, Garching, Germany
[2] LESIA, Observatoire de Paris, Université PSL, Sorbonne Université, Université Paris Cité, CNRS, Meudon, France
[3] Max Planck Institute for Astronomy, Heidelberg, Germany
[4] 1st Institute of Physics, University of Cologne, Cologne, Germany
[5] Université Grenoble Alpes, CNRS, IPAG, France
[6] Faculty of Sciences, University of Lisbon, Portugal
[7] Faculty of Engineering, University of Porto, Portugal
[8] ESO
[9] Leiden Observatory, Leiden University, the Netherlands
[10] Departments of Physics and Astronomy, University of California, Berkeley, USA
[11] CENTRA – Centre for Astrophysics and Gravitation, University of Lisbon, Portugal
[12] Department of Physics, Technical University of Munich, Germany
[13] Max Planck Institute for Radio Astronomy, Bonn, Germany
[14] Hamburg Observatory, University of Hamburg, Germany
[15] School of Physics & Astronomy, University of Southampton, UK
[16] Institute of Astronomy, KU Leuven, Belgium
[17] Institute of Astronomy, Cambridge, UK
[18] Research School of Astronomy and Astrophysics, Australian National University, Canberra, Australia

[19] Université Côte d'Azur, Laboratoire Lagrange, France
[20] Université Lyon 1, ENS de Lyon, CNRS, Lyon, France

The GRAVITY instrument has been revolutionary for near-infrared interferometry by pushing sensitivity and precision to previously unknown limits. With the upgrade of GRAVITY and the Very Large Telescope Interferometer (VLTI) in GRAVITY+, these limits will be pushed even further, with vastly improved sky coverage, as well as faint-science and high-contrast capabilities. This upgrade includes the implementation of wide-field off-axis fringe-tracking, new adaptive optics systems on all Unit Telescopes, and laser guide stars in an upgraded facility. GRAVITY+ will open up the sky to the measurement of black hole masses across cosmic time in hundreds of active galactic nuclei, use the faint stars in the Galactic centre to probe General Relativity, and enable the characterisation of dozens of young exoplanets to study their formation, bearing the promise of another scientific revolution to come at the VLTI.

## Introduction

The near-infrared interferometric beam-combiner GRAVITY has been in operation at the Very Large Telescope Interferometer (VLTI) for five years (GRAVITY Collaboration et al., 2017). During that time GRAVITY has transformed optical interferometry by delivering ground-breaking results in studies of the Galactic centre, active galactic nuclei (AGN), exoplanets, and young stellar objects (see for example GRAVITY Collaboration et al., 2018a,b,c; GRAVITY Collaboration et al., 2019a,b). GRAVITY can achieve microarcsecond astrometry and detect stars as faint as 20th *K*-band magnitude in the Galactic centre. The current performance is not the ultimate within reach, which motivated the upgrade of the instrument to GRAVITY+. Studies of the performance of the VLTI also highlighted the importance of infrastructure upgrades to increase the overall sensitivity (Mérand, 2018).

Following this, GRAVITY+ was first presented in 2019, and after a phase A study in 2021 the ESO Council approved the project in December 2021. Since then the phased implementation of GRAVITY+ has already started and is combining upgrades of the GRAVITY instrument and of the VLTI infrastructure itself. In its final form, GRAVITY+ will provide much better sensitivity, by adding state-of-the-art adaptive optics (AO) systems and laser guide stars (LGS) to all four Unit Telescopes (UTs) of the VLTI, and increase the field for picking a phase reference star from the current 2 arcseconds to 30 arcseconds for significantly improved sky coverage. Most of the upgrades serve all current and future VLTI instruments, in addition to GRAVITY itself.

## Upgrades

### Adaptive optics and laser guide stars upgrade

The need for a bright AO guide star and the limited performance of the current MACAO system on the UTs strongly limit the sky coverage with AO, and thereby interferometry at the VLTI. This limitation is addressed in GRAVITY+ by adding LGS and a new AO system to be installed on all UTs. GRAVITY+ will equip UT1, 2, and 3 with a side-launch LGS, the design of which is shared with the ESO's Extremely Large Telescope's LGS, and use one of the already existing Adaptive Optics Facility 4LGSF on UT4 (see Figure 1). The new state-of-the-art AO includes new wavefront sensors equipped with a natural guide star (NGS) and an LGS module each, with a high-order Shack-Hartmann sensor and a 1353-actuator deformable mirror. These new AO systems will dramatically increase the Strehl ratios and thereby the flux injection into the optical fibres of GRAVITY. Especially for faint targets, the improved AO will yield a flux increase in the fibre by up to a factor of 10. Furthermore, the new AO system will improve operability by reducing the demand for the very best atmospheric conditions for faint interferometry. In NGS operation the expected Strehl ratio for bright objects is greater than 80 % in the *K* band, and in LGS operation the expected limiting magnitude is $R$ = 18 mag with a *K*-band Strehl of more than 40 %. With the improved AO performance, the limiting magnitude of the fringe-tracking star improves from currently $K$ ~ 10 mag to $K$ = 13 mag. The probability of finding such a fringe-tracking star within 30 arcseconds is almost 100% in the Galactic plane and remains above 25% for Galactic latitudes of 40° (Figure 2).

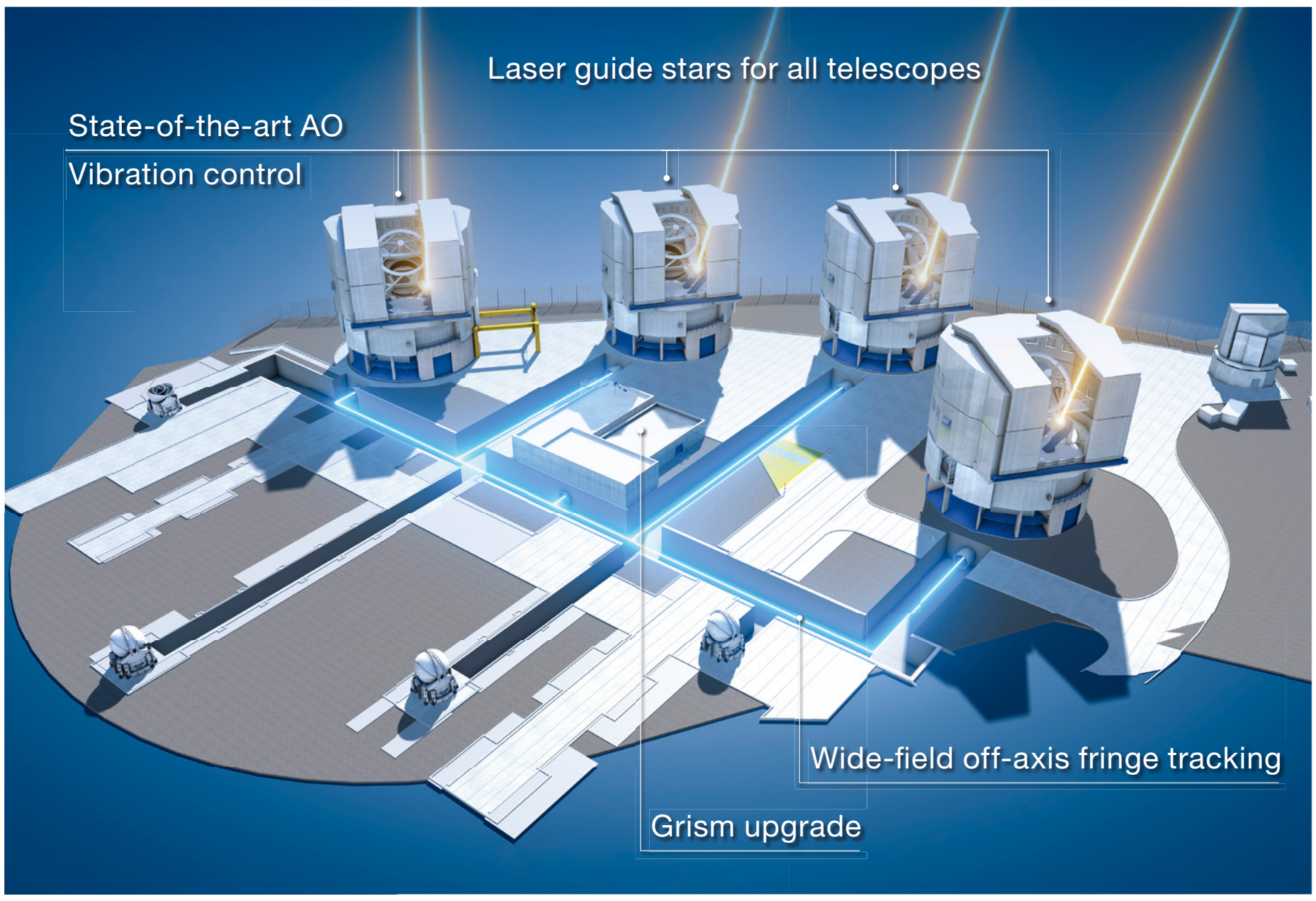

Figure 1. Overview of all GRAVITY+ components.

### Off-axis fringe tracking

GRAVITY's highest sensitivity is achieved in the dual-beam mode, in which a nearby star is used for fringe tracking instead of the science target itself. In GRAVITY the separation between the science object and the fringe-tracking star has to be smaller than 2 arcseconds for the UTs and 4 arcseconds for the ATs, which significantly limits the number of observable targets. To overcome this limitation, one part of GRAVITY+ is the new off-axis fringe-tracking mode, called GRAVITY-Wide (GRAVITY+ Collaboration et al. 2022). In this mode, two subfields of the telescope field of view are selected by the star separators located at the coudé foci of the telescopes. One of the two fields contains the science source and the other one the fringe-tracking star. The two fields are brought separately into the VLTI lab and then both fields are fed into the GRAVITY beam-combiner. This mode allows much larger separations between the two objects than were previously possible. Separations of up to several tens of arcseconds are possible, effectively limited only by the coherence loss in the atmosphere, induced by the differential piston. GRAVITY-Wide is already fully commissioned and available to the community (GRAVITY+ Collaboration et al. 2022). Together with the LGS system this new mode significantly improves the sky coverage, as shown in Figure 2.

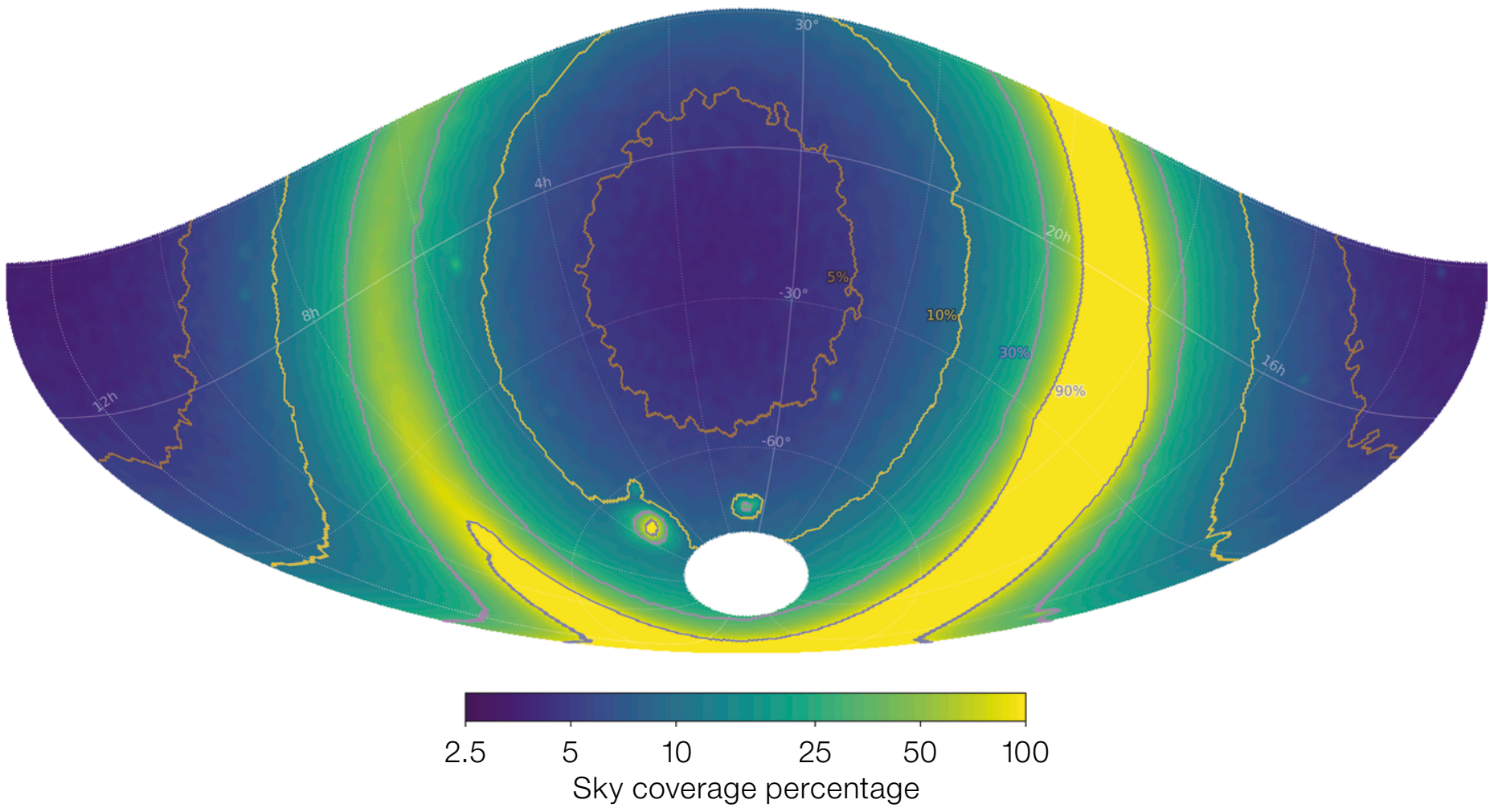

Figure 2. Sky coverage for LGS AO-supported off-axis fringe tracking with a fringe-tracking star as faint as $m_K = 13$, and a maximum allowed separation of 30 arcseconds. The sky projection is centred on the zenith in the Chilean spring. Areas not observable by the VLTI are left blank. This sky coverage is orders of magnitude larger than the current capability of the VLTI.

### Sensitivity

GRAVITY+ also encompasses several other projects to increase the sensitivity of the instrument and reduce existing noise sources. In October 2019 two grisms were replaced in GRAVITY's science spectrometer. This upgrade yielded an improvement of a factor of 2–3 in throughput in the medium- and high-resolution modes (Yazici et al., 2021).

Another goal of this project is to reduce optical path differences (OPDs) in the VLTI. These OPDs come mainly from vibrations in the telescopes and affect the performance of the fringe tracker in GRAVITY. The upgrade of the vibration control system MANHATTAN2 is currently being commissioned at Paranal, to measure and compensate for existing high-frequency vibrations. Together with a new implementation of the fringe tracker, this will lead to a much improved fringe-tracking performance.

Finally, one of the main noise sources in GRAVITY is the back-fluorescence of the metrology laser into the spectrometer, which originates in the optical fibres of GRAVITY. Within GRAVITY+, we are developing new observing modes for the faintest targets in which the noise from the metrology laser is removed without losing the stability of the instrument (Widmann et al., 2022), by toggling the brightness of the laser beams between exposure times and presets.

### Phased Implementation

The implementation of GRAVITY+ encompases three phases. The first phase is almost concluded, with the upgrade of the GRAVITY grisms and the implementation of the off-axis fringe-tracking mode already finished. Other parts of this phase, such as the improved fringe tracker implementation and the reduction of vibrations, are currently ongoing. The second phase of the project will see the replacement of the MACAO AO system with the GRAVITY+ AO system. In the last phase, which will conclude the GRAVITY+ implementation, the three remaining UTs will be equipped with LGS. The full project is expected to be completed and available to the community in 2026. The phased approach has the advantage that the upgrades have minimal impact on the normal VLTI operation while adding new capabilities at each step.

## Science goals

### AGN at high redshift

Supermassive black holes (SMBHs) are expected to be present at the centres of all massive galaxies in the Universe. The properties of the black holes are tightly correlated with the properties and evolution of their host galaxy. Crucial to understanding this co-evolution of the SMBHs and their galaxies is knowing the masses of the black holes. Especially for AGN, one cannot measure black hole masses directly but has to use indirect methods, combining velocity information from the AGN spectra with size information typically based on scaling relations that are calibrated via reverberation mapping (Peterson et al., 2004). However, spectroastrometry with GRAVITY can directly measure the sizes of the broad-line regions around SMBHs (GRAVITY Collaboration et al. 2018c). From the size measurement of the broad-line region, one can directly infer the mass of the black hole. The sample of AGN currently within reach of GRAVITY is limited by the ability to fringe track on the AGN itself and the performance of the AO, leaving only a few AGN observable. With up to 30-arcsecond separations for external fringe tracking and significantly improved sensitivity, GRAVITY+ will increase the number of accessible AGN to a few 100, making it a true cosmic explorer (see Figure 3).

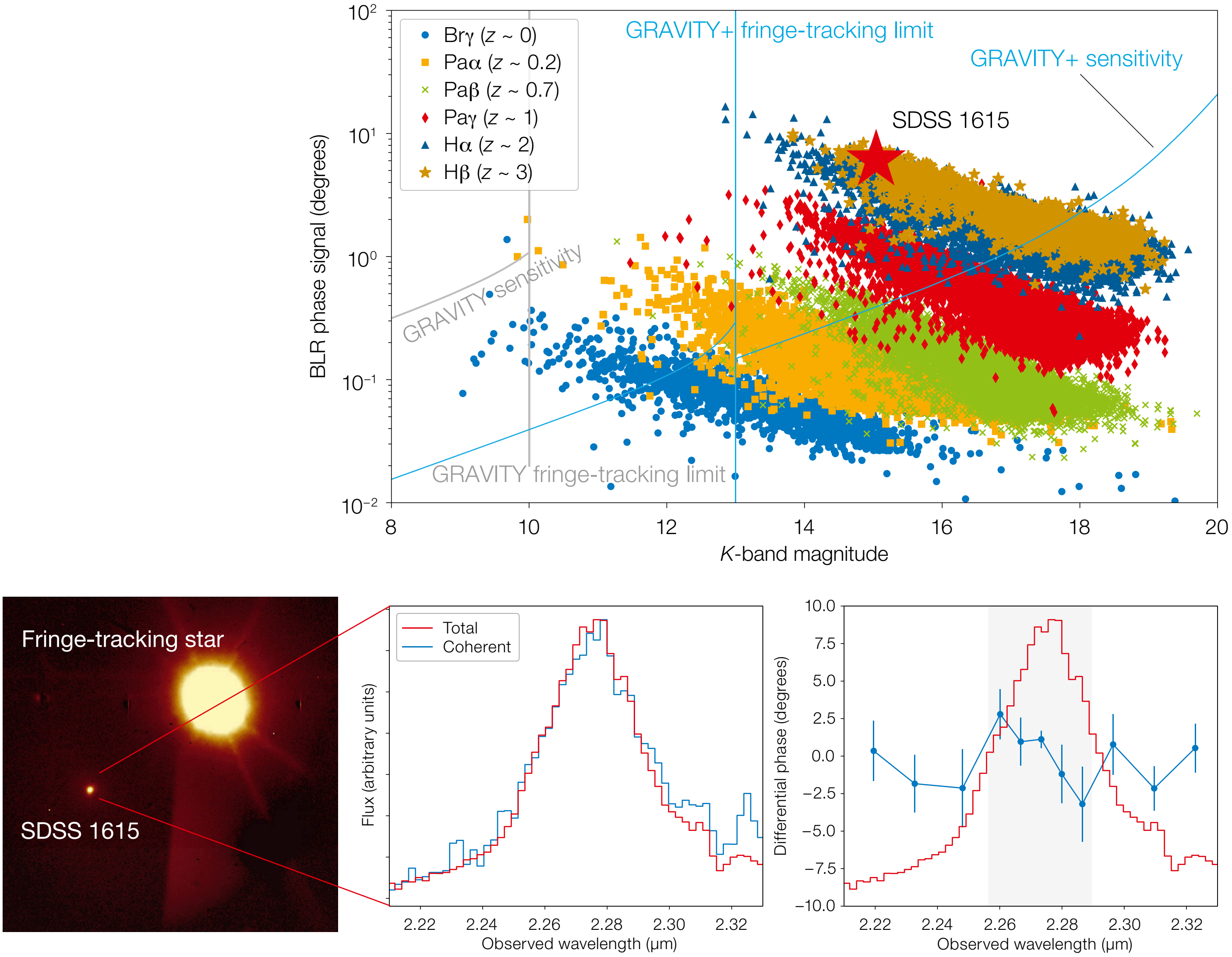

**Figure 3.** Top: Observable AGN with GRAVITY+. At different redshifts, different spectral lines are observable in the *K* band, indicated by different colours. The observing limits in terms of fringe-tracking brightness (vertical line) and sensitivity (diagonal line) are shown for GRAVITY in grey and GRAVITY+ in blue. Targets fainter than the fringe-tracking limit will still be observable via GRAVITY-Wide. Bottom: The first successful observation of a high-redshift AGN with near-infrared interferometry. The left panel shows the target on the acquisition camera, which is indicated as an asterisk in the top plot. The right two panels show the coherent flux (middle) and the differential phase signal (right) from the AGN.

Measurements of SMBH masses around $z = 2$ are especially interesting as this was the peak of star formation in the Universe, which makes it a crucial time for the co-evolution of galaxies and their SMBHs. With the commissioning of GRAVITY-Wide, we could for the first time demonstrate the ability to observe a $z > 2$ AGN. As shown in Figure 3 the coherent flux of the AGN is clearly detected. This target was observed for only a little more than one hour, but we could already extract a tentative signal in the differential phase (for more information, see GRAVITY+ Collaboration et al. 2022).

## The Galactic centre

By observing stars orbiting the SMBH Sgr A* in the Galactic centre, GRAVITY has delivered precision tests of Einstein's general theory of relativity. In the orbit of the star S2, the effects of the gravitational redshift and the Schwarzschild precession have been observed (GRAVITY Collaboration et al. 2018a & 2020a). Together with the observation of material orbiting the central source close to the innermost stable orbit (GRAVITY Collaboration et al. 2018b), these observations have delivered the strongest evidence to date that Sgr A* is indeed a Schwarzschild-Kerr black hole. With its increased sensitivity and Strehl ratio GRAVITY+ will open up new possibilities in the Galactic centre. One is the search for fainter stars on close orbits around Sgr A*. The faintest star found with GRAVITY so far has a magnitude of $K > 19$ mag (GRAVITY Collaboration et al. 2022, and see Figure 4), but even fainter stars are expected to surround Sgr A*. Such a faint star on a close orbit would allow for the first time the measurement of the effects of the black hole's spin in a stellar orbit. A measurement of the spin would help to understand the accretion physics of massive black holes in a main-sequence disc galaxy and the interplay between Sgr A* and the accretion flow surrounding it. GRAVITY+ will be crucial for getting the astrometric accuracy needed to measure the effect of the spin on a star in the close neighbourhood of Sgr A* (see Figure 4). These observations will be complementary to the spectroscopic measurements with ERIS on the VLT and MICADO on the ESO's Extremely Large Telescope.

Not only will the detection of fainter stars be possible with GRAVITY+, but also the observations of flare motions and their polarisation properties will vastly improve. This will allow a better understanding of the magnetic field structure and hot gas motion on the innermost stable orbit around Sgr A* (GRAVITY Collaboration et al. 2018b). This would, for example, allow a test of whether for ultra-low-accreting black holes the spin of the black hole and the angular momentum vector of the accretion flow align via the Bardeen-Petterson effect (Bardeen & Petterson, 1975). Current simulations suggest that this should not be the case (Ressler, Quataert & Stone, 2018).

## Exoplanets

The unique observing capabilities of GRAVITY have delivered the first characterisation of an exoplanet atmosphere in the *K* band with interferometry (GRAVITY Collaboration et al. 2019a), and astrometry with better than 50-microarcsecond accuracy, a factor of 50–100 more precise than conventional imaging techniques. These observations allow for the characterisation of young planets in the ice-line region at 2–3 au orbital separation from the host star, probing the location where these planets form (GRAVITY Collaboration et al. 2020b; Wang et al., 2021). The dramatic increase of AO performance with the high-order AO of GRAVITY+ will extend the inner working angle 5–8 times closer to the star than traditional direct-imaging instruments. The upgraded performance of GRAVITY+ will open up new parameter space for the exoplanet population, increasing the sample of directly imaged exoplanets by an order of magnitude. Precise host-star astrometry with GAIA will be a prime technique for determining prior estimates of where GRAVITY can search for planets, generating a large sample of observable planets in the 1–2 au region. This was recently demonstrated for HD 206893 c (Hinkley et al., 2022). The simultaneous measurement of luminosity with GRAVITY+ and masses with GAIA will enable the measurement of the mass-luminosity relation for dozens of planets, and thus constrain the initial entropy of these objects, which is key for understanding their formation. In addition, GRAVITY+ will be able to measure the atmospheric properties of gas-giant planets with significantly higher sensitivity, particularly the C/O ratio, which constrains the location and time of formation for these planets. Finally, the astrometric precision of GRAVITY+ will provide measurements of orbital architectures and dynamics of exoplanetary systems with unmatched precision, even in the era of extremely large telescopes, as already demonstrated with GRAVITY (Lacour et al., 2021).

## Further science cases

In addition to the science cases mentioned here, the increased capabilities of GRAVITY+ will benefit many other observations. The increased sensitivity and sky-coverage of GRAVITY+ will enable the observation of embedded, low-mass young stellar objects at the onset of planet formation. GRAVITY+ will be able to trace accreted and ejected gas, spatially resolved at a few stellar radii (Gravity Collaboration et al. 2020c). GRAVITY was also used to characterise compact objects via microlensing (see, for example, Dong et al., 2019). While only a few such events are within reach for GRAVITY, the upgrade will give access to thousands of microlensing events. Similarly, GRAVITY+ will permit resolving massive stars and searching for intermediate-mass black holes in globular clusters from precise tracking of stellar motions. The variety of different fields shows that GRAVITY+ is indeed an upgrade that will enable many new and unique science cases.

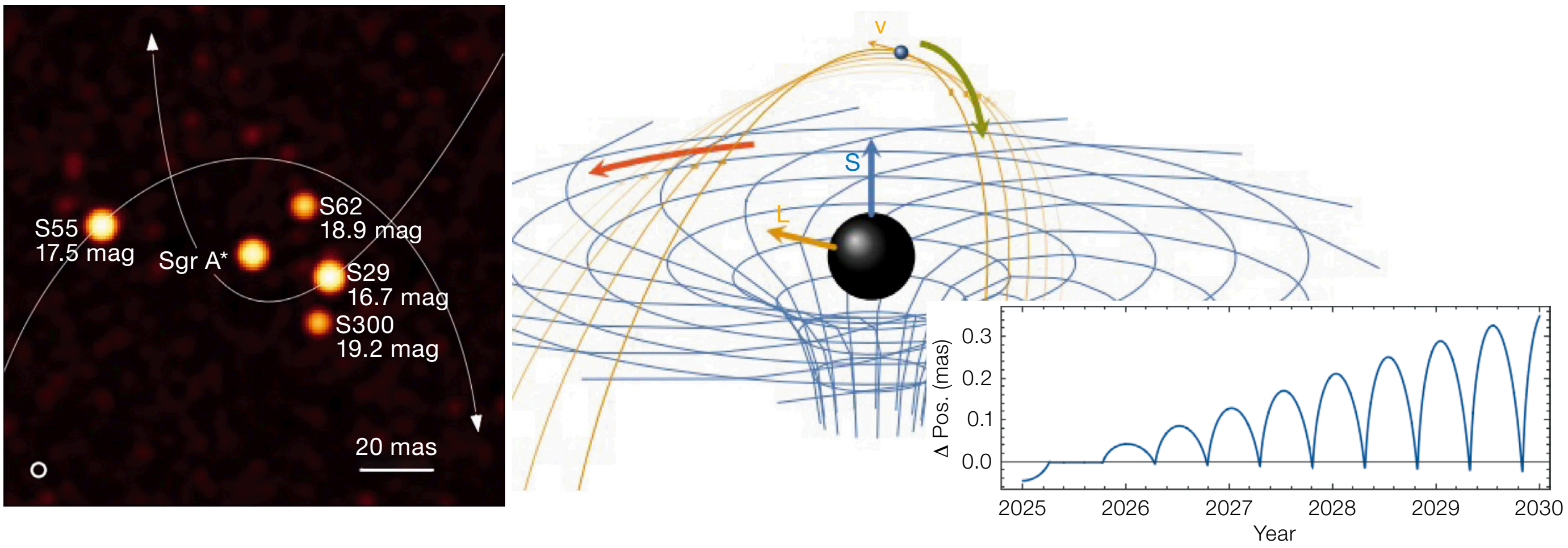

Figure 4. Left: Image of the stars around Sgr A* taken with GRAVITY. Right: Illustration of the Lense-Thiring precession of a star caused by a spinning black hole. The insert shows the effect on a hypothetical star with a semi-major axis ten times smaller than that of S2.

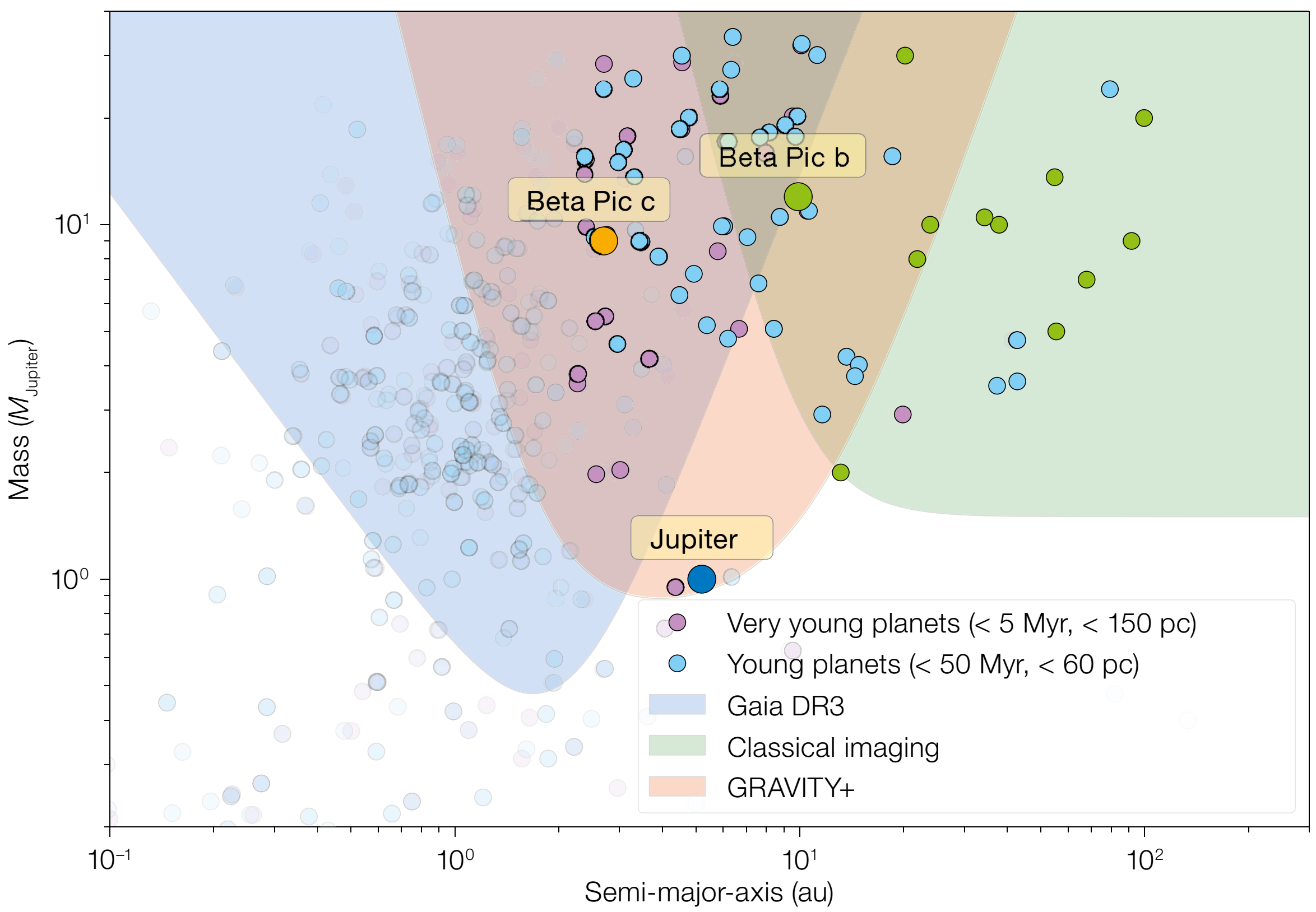

**Figure 5.** Exoplanet population as a function of semi-major axes and masses. GRAVITY+ will give access to a new part of the parameter space for young planets in the 1–10 au range. Potential GRAVITY+ detections are indicated by solid points.

## Conclusion

The GRAVITY instrument has transformed near-infrared interferometry. GRAVITY+ will continue to reshape this field with a major upgrade of the GRAVITY instrument and the VLTI infrastructure, thus also serving other VLTI instruments. The biggest leap will come with the implementation of a state-of-the-art AO system with LGS. GRAVITY will greatly benefit on many fronts from the improved AO system as it ensures a more stable flux injection into the optical fibres, which means a more stable fringe-tracker performance and a higher sensitivity for fainter targets. Together with the off-axis fringe-tracking mode, this will lead to a dramatic increase in sky coverage for GRAVITY+. With all the improvements GRAVITY+ will show an overall improvement in performance of 4–5 magnitudes and push the limiting magnitude down to approximately $K = 22$ mag. These performance improvements will lead to new and unique scientific opportunities. With the observation of galaxy and SMBH co-evolution around cosmic noon, the possible measurement of the spin of a SMBH, and the detection and characterisation of exoplanets in an otherwise unprobed regime, GRAVITY+ will stay at the frontier of astronomy. The combination of 8-metre-class telescopes at the VLTI and a highly sensitive interferometer itself is unmatched in the world. GRAVITY+, with its unique and timely scientific opportunities, ensures that the VLTI will deliver otherwise impossible scientific results and that it remains unique, even in the era of the 30–40-metre-class telescopes.

### References

Bardeen, J. M. & Petterson, J. A. 1975, ApJ, 195, L65
Dong, S. et al. 2019, ApJ, 871, 70
GRAVITY Collaboration et al. 2017, A&A, 602, A94
GRAVITY Collaboration et al. 2018a, A&A, 615, L15
GRAVITY Collaboration et al. 2018b, A&A, 618, L10
GRAVITY Collaboration et al. 2018c, Nature, 563, 657
GRAVITY Collaboration et al. 2019a, A&A, 623, L11
GRAVITY Collaboration et al. 2019b, A&A, 632, A53
GRAVITY Collaboration et al. 2020a, A&A, 636, L5
GRAVITY Collaboration et al. 2020b, A&A, 633, A110
GRAVITY Collaboration et al. 2020c, Nature, 584, 547
GRAVITY Collaboration et al. 2022, A&A, 657, A82
GRAVITY+ Collaboration et al. 2022, A&A, 665, A75
Hinkley, S. et al. 2022, submitted to A&A
Lacour, S. et al. 2021, A&A, 654, L2
Mérand, A. 2018, The Messenger, 171, 14
Peterson, B. M. et al. 2004, ApJ, 613, 682
Ressler, S. M., Quataert, E. & Stone, J. M. 2018, MNRAS, 478, 3544
Yazıcı, Ş. et al. 2021, Proc. SPIE, 11446, 114461X
Wang, J. J. et al. 2021, AJ, 161, 148
Widmann, F. et al. 2022, Proc. SPIE, 12183, 121830U